\documentclass[epj,twocolumn]{webofc}
\usepackage[varg]{txfonts}   % Web of Conferences font
\usepackage{epstopdf}

\woctitle{The Innermost Regions of Relativistic Jets and Their Magnetic Fields}

\begin{document}

\title{Analyzing polarization swings in 3C\,279}

\author{
	S.~Kiehlmann\inst{1}\fnsep\thanks{\email{skiehlmann@mpifr-bonn.mpg.de}}
	\and
	T.~Savolainen\inst{1}
	\and
	S.~G.~Jorstad\inst{2,3}
        \and
        K.~V.~Sokolovsky\inst{4,11}        
        \and 
        F.~K.~Schinzel\inst{5}
        \and
	I.~Agudo\inst{6,2,7}
	\and
	A.~A.~Arkharov\inst{8}
        \and 
        E.~Ben\'itez\inst{9}
        \and
        A.~Berdyugin\inst{10}
	\and
	D.~A.~Blinov\inst{3}
	\and
        N.~G.~Bochkarev\inst{11}
        \and
        G.~A.~Borman\inst{12}
        \and
        A.~N.~Burenkov\inst{13}
        \and
	C.~Casadio\inst{6}
	\and
        V.~T.~Doroshenko\inst{14,12}
        \and
	N.~V.~Efimova\inst{3}
        \and
        Y.~Fukazawa\inst{15}
        \and
        J.~L.~G\'omez\inst{6}
	\and
	V.~A.~Hagen-Thorn\inst{3,16}
        \and
        J.~Heidt\inst{17}
	\and
        D.~Hiriart\inst{18}
        \and
        R.~Itoh\inst{15}
        \and
	M.~Joshi\inst{2}
	\and
        G.~N.~Kimeridze\inst{19}
        \and
	T.~S.~Konstantinova\inst{3}
	\and
	E.~N.~Kopatskaya\inst{3}
        \and
        I.~V.~Korobtsev\inst{20}
        \and
        Y.~Y.~Kovalev\inst{4,1}
        \and
        T.~Krajci\inst{21}
	\and
        O.~Kurtanidze\inst{19,17}
        \and
        S.~O.~Kurtanidze\inst{19}
        \and
	V.~M.~Larionov\inst{3,8,16}
	\and
	E.~G.~Larionova\inst{3}
	\and
	L.~V.~Larionova\inst{3}
        \and
        E.~Lindfors\inst{22}
        \and
        J.~M.~L\'opez\inst{18}
	\and
	A.~P.~Marscher\inst{2}
	\and
	I.~M.~McHardy\inst{23}
	\and
	S.~N.~Molina\inst{6}
	\and
	D.~A.~Morozova\inst{3}
	\and
        S.V.~Nazarov\inst{12}
        \and
        M.~G.~Nikolashvili\inst{19}
        \and
        K.~Nilsson\inst{22}
        \and
        N.~G.~Pulatova\inst{12}
        \and
        R.~Reinthal\inst{10}
        \and
        A.~Sadun\inst{24}
        \and
        S.~G.~Sergeev\inst{12}
        \and
        L.~A.~Sigua\inst{19}
        \and
        M.~Sorcia\inst{9}
        \and
        O.~I.~Spiridonova\inst{13}
        \and
        L.~O.~Takalo\inst{10}
        \and
	B.~Taylor\inst{2,25}
	\and
	I.~S.~Troitsky\inst{3}
        \and
        L.~S.~Ugolkova\inst{11}
        \and
        J.~A.~Zensus\inst{1}
        \and
        V.~E.~Zhdanova\inst{13} 
	}

\institute{
	Max-Planck-Institut f\"ur Radioastronomie, Auf dem H\"ugel, 69,
        D-53121, Bonn, Germany
	\label{1}
	\and
	Institute for Astrophysical Research, Boston University, 
        725 Commonwealth Avenue, Boston, MA 02215, USA
	\label{2}
        \and
	Astronomical Institute, St. Petersburg State University, 
        Universitetskij Pr. 28, Petrodvorets, 198504 St. Petersburg, Russia
	\label{3}
	\and
        Astro Space Center of Lebedev Physical Institute, Profsoyuznaya 84/32, 
        117997, Moscow, Russia
        \label{4}
        \and
        Department of Physics and Astronomy, University of New Mexico, 
        Albuquerque NM, 87131, USA 
        \label{5}
        \and
	Instituto de Astrof\'{\i}sica de Andaluc\'{\i}a, CSIC, Apartado 3004, 
        18080, Granada, Spain
	\label{6}
	\and
        Joint Institute for VLBI in Europe, Postbus 2, 7990 AA, Dwingeloo, 
        the Netherlands
        \label{7}
        \and
	Main (Pulkovo) Astronomical Observatory of RAS, Pulkovskoye shosse, 60, 
        196140, St. Petersburg, Russia
	\label{8}
	\and
        Instituto de Astronom\'ia, Universidad Nacional Aut\'onoma de M\'exico,
        04510 M\'exico DF, M\'exico
        \label{9}
        \and
        Tuorla Observatory, Department of Physics and Astronomy, University of Turku, 
        V\"ais\"al\"antie 20, 21500, Piikki\"o, Finland
        \label{10}
        \and
        Sternberg Astronomical Institute, M.V.Lomonosov Moscow State University, 
        Universiteskij prosp. 13, Moscow 119991, Russia
	\label{11}
        \and
        Crimean Astrophysical Observatory, P/O Nauchny Crimea 98409, Ukraine 
	\label{12}
        \and
        Special Astrophysical Observatory of the Russian AS, Nizhnij Arkhyz, 
        Karachaevo-Cherkesia 369167, Russia 
        \label{13}
        \and
        Southern station of the Moscow Lomonosov State University, Moscow, Russia,
        P/O Nauchny, 98409 Crimea, Ukraine 
	\label{14}
        \and 
        Department of Physical Sciences, Hiroshima University, Higashi-Hiroshima, 
        Hiroshima 739-8526, Japan
        \label{15}
        \and
	Isaac Newton Institute of Chile, St. Petersburg Branch, St. Petersburg, Russia
	\label{16}
	\and
        ZAH, Landessternwarte Heidelberg, K\"onigstuhl 12, 69117 Heidelberg, Germany
        \label{17}
        \and
        Instituto de Astronom\'ia, Universidad Nacional Aut\'onoma de M\'exico,
        22860 Ensenada BC, M\'exico
        \label{18}
        \and
        Abastumani Observatory, Mt. Kanobili, 0301 Abastumani, Georgia
        \label{19}
        \and
        Inst. of Solar-Terrestrial Physics, Lermontov st. 126a, Irkutsk p/o box 291, 
        664033, Russia
	\label{20}
        \and
        Center for Backyard Astrophysics - New Mexico, PO Box 1351 Cloudcroft, 
        NM 88317, USA
        \label{21}
        \and
        Finnish Centre for Astronomy with ESO (FINCA), University of Turku, 
        V\"ais\"al\"antie 20, FI-21500 Piikki\"o, Finland
	\label{22}
	\and
	Department of Physics and Astronomy, University of Southampton, 
        Southampton, SO17 1BJ, United Kingdom
        \label{23}
        \and
        Department of Physics, University of Colorado Denver, CO, USA 
        \label{24}
        \and
	Lowell Observatory, Flagstaff, AZ 86001, USA
	\label{25}
}

\abstract{ Quasar 3C\,279 is known to exhibit episodes of optical
  polarization angle rotation. We present new, well-sampled optical
  polarization data for 3C\,279 and introduce a method to distinguish
  between random and deterministic electric vector position angle
  (EVPA) variations. We observe EVPA rotations in both directions with
  different amplitudes and find that the EVPA variation shows
  characteristics of both random and deterministic cases. Our analysis
  indicates that the EVPA variation is likely dominated by a random
  process in the low brightness state of the jet and by a
  deterministic process in the flaring state. }

\maketitle

%_NEW_SECTION___________________________________________________________________
\section{Introduction}
\label{sec:intro}

Rotations of the electric vector position angle (EVPA) of linearly
polarized radiation have been observed in various blazars since the
1960s \cite[e.g.,][]{1967ApJ...148L..53K}.  Proposed explanations
include stochastic variation \cite[e.g.,][]{1985ApJ...290..627J},
multi-component models \cite[cf.][]{1984MNRAS.211..497H}, and bent
trajectories of a moving shock in a tangled
\cite[cf.][]{1990ApJ...350..536C} or structured
\cite[cf.][]{1988A&A...190L...8K} magnetic field. In 3C\,279 a
$300\,^\circ$ counter-clockwise rotation of the optical EVPA was
observed in 2006/2007 and explained by an emission feature on a spiral
path in a helical magnetic field \cite{2008A&A...492..389L}.  In 2009
a $210\,^\circ$ rotation in the opposite direction was observed and
explained by a global bend of the jet \cite{2010Natur.463..919A}.

A major issue in analyzing polarization data is the $n \pi$ ambiguity
of the measured EVPA.  Low sampling rates may obscure actual, large
rotations and complicate the reconstruction and interpretation of the
actual EVPA variation.  We discuss here a method to distinguish random
from deterministic EVPA variation and apply this method to our data
set of optical polarization observations of 3C\,279 from November 2010
to August 2012.

%_NEW_SECTION___________________________________________________________________
\section{Data set and data processing}
\label{sec:data}

\begin{table}
\centering
\caption{Observatories participating in optical polarimetric monitoring.}
\label{tab:poldata}
{\small
\begin{tabular*}{\columnwidth}{@{\extracolsep{\fill} } lrl}
\hline \hline
	Institution		& Tel. diam.	& Filters \\
	\hline
%	\multicolumn{3}{c}{\textit{optical}} \\
	Calar Alto, Spain	        & 220\,cm	& R \\
	CrAO, Ukraine		        & 70\,cm	& R \\
	Kanata, Japan		        & 150\,cm	& V \\
	KVA, La Palma, Spain	        & 60\,cm	& white light \\
	Liverpool, La Palma, Spain      & 200\,cm	& V+R \\
	Perkins, USA		        & 183\,cm	& R \\
	San Pedro M\'artir, Mexico	& 84\,cm	& R \\
	St. Petersburg, Russia          & 40\,cm	& white light \\
	Steward Observatory, USA	& 154 and 229\,cm	& 5000-7000\AA \\
\hline
\end{tabular*}
}
\end{table}

The photometric R-band data set consists of 24 light curves provided
by different observers and institutions.  We combine these individual
light curves into one assembled light curve after removing clear
outlier data points, averaging intra-night data, and
cross-calibrating the individual light curves. Optical polarimetry
data are provided by nine observatories listed in
Table~\ref{tab:poldata}.  Since frequency-dependence of the EVPA in
optical is relatively weak, we combine all optical EVPA data sets to
solve for the $n \pi$-ambiguity.

%_NEW_SUBSECTION________________________________________________________________
\subsection{EVPA ambiguity}
\label{sec:evpaambiguity}

Comparison of EVPA measurements is ambiguous.  The measured properties
are the same for an EVPA $\chi$ and the EVPA $\chi' = \chi \pm n \cdot
\pi$ with $n \in \mathbb{N}$; thus, EVPA rotations larger than $\pi$
cannot be determined unambiguously. The common procedure to solve for
this ambiguity is to assume a smooth variation of the EVPA and to
shift data points to minimize the difference between adjacent data
points (see e.g., \cite{2010Natur.463..919A}). We employ two methods
to shift EVPA data points:

\begin{description}
	\item[EVPA shifting method 1] determines the median of the $N$
          previous data points as a reference for an EVPA data point
          and calculates the absolute deviation between them.
	\item[EVPA shifting method 2] calculates the absolute
          deviation between two adjacent data points and subtracts the
          root summed squared errors:
	\begin{align}
		\Delta \chi_\mathrm{est} = \left| \chi_i - \chi_{i-1} \right| - \sqrt{\sigma^2 (\chi_i) + \sigma^2 (\chi_{i-1})}.
	\end{align}
\end{description}

Both methods shift an EVPA data point by $\pm n \pi$ if the deviation
is larger than $\pi/2$ with $n \in \mathbb{N}$ such that the deviation
is minimized. The first method with large $N$ is robust against
measurement errors, but it is likely to obscure real variation, if the
data is not well sampled.  With our worst data sampling of one data
point each 20 days, $N=2$ must not be exceeded to reconstruct EVPA
rotations rates up to $3.6\,^\circ/\mathrm{d}$ (measured with
method~2).  The fast rotations are sampled much better; the mean
sampling rate allows for $N=16$.  We choose $N=4$. The second method
is expected to obscure real variation less frequently than
method~1. With a sampling rate and errors similar to those of our
observed data, both methods correctly reconstruct a simulated EVPA
rotation of $500\,^\circ / 200\,\mathrm{d}$ with a probability of
$98.5\,\%$.

%_NEW_SUBSECTION________________________________________________________________
\subsection{Optical light curve and polarization}
\label{sec:pollightcurve}

\begin{figure}
\centering
\includegraphics[width=\columnwidth]{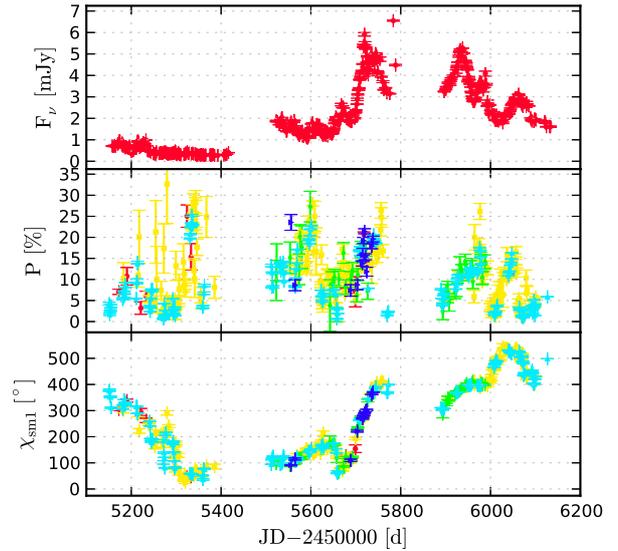}
\caption{Optical polarization light curves of 3C\,279.  \textit{Upper
    panel:} R-band flux density.  \textit{Mid panel:} Linear
  polarization fraction at V-band (red circles), R-band (yellow
  squares), V+R (green diamonds), 5000-7000\,\AA\ (pale blue up-sided
  triangles), white light (dark blue right-sided
  triangles). \textit{Lower panel:} EVPA shifted with method~1
  (symbols as in the second panel).}
\label{fig:pollightcurve}
\end{figure}

Figure~\ref{fig:pollightcurve} shows the assembled R-band light curve,
the degree of optical linear polarization, $P$, and the optical EVPA
shifted with method~1, $\chi_\mathrm{sm1}$.  The optical light curve
shows little variation and low flux density, $F_\nu <
1\,\mathrm{mJy}$, before $\mathrm{JD} = 2455400$.  After the first
seasonal gap the light curve covers a flaring state with an overall
flux increase, two major flares, several smaller flares, and a flux
decrease.  During the first optical flare the flux density increases
by a factor of 3 within 30~days.  The mean linear polarization degree
is $\left< P \right> = 12\,\%$ with a standard deviation $\sigma(P) =
8\,\%$.  Shifting the EVPA with method~1, we observe EVPA rotations in
both directions with amplitudes up to $360\,^\circ$.  A smooth
$360\,^\circ$-rotation coincides with the first major optical
flare. To test whether these EVPA rotations are of a random or a
deterministic origin we use the scheme presented in the following
section.

%_NEW_SECTION___________________________________________________________________
\section{Discriminating random from deterministic EVPA variation}
\label{sec:randomEVPA}

Shifting the EVPA data points to solve the $n\pi$-ambiguity is based
on the assumption that the EVPA actually rotates smoothly rather than
jumps abruptly. This assumption leads to two questions.  First, is it
valid?  If the underlying process is of a random origin, shifting EVPA
data points artificially introduces rotations. And second, what does
\emph{smooth} mean for discretely sampled measurements?
Jones~et~al.~\cite{1985ApJ...290..627J} have shown that random
variations of the EVPA can produce rotations $>360\,^\circ$.

We define a quantitative measure of the \emph{smoothness} of an EVPA
curve and define a random process based on \cite{1985ApJ...290..627J}
and \cite{2007ApJ...659L.107D} to determine the probability of a
random EVPA curve to show a large ($>180\,^\circ$) and smooth
rotation.

\begin{table}[t]
\centering
\caption{Simulation input parameters.}
\label{tab:simparameters}
{\small
\begin{tabular*}{\columnwidth}{@{\extracolsep{\fill} } l l r}
	\hline \hline
	time interval	& $T$		& $200\,\mathrm{d}$	\\
	mean time step	& $\left< \Delta t \right>$	& $3\,\mathrm{d}$	\\
	min. time step:	& $\Delta t_\mathrm{min}$	& $0.7\,\mathrm{d}$	\\
	max. time step:	& $\Delta t_\mathrm{max}$	& $21\,\mathrm{d}$	\\
	$\Delta t$ power-law index:	& $\alpha$	& $-2$	\\
	error level:	& $\sigma_\mathrm{errors}$	& $4\,^\circ$	\\
	\hline
\end{tabular*}
}
\end{table}

%_NEW_SUBSECTION________________________________________________________________
\subsection{A quantitative measure of smoothness}
\label{sec:smoothness}

The point-to-point variation of the EVPA curve is
\begin{align}
  \left( \frac{\Delta \chi}{\Delta t} \right)_i = \frac{\chi_i - \chi_{i-1}}{t_i - t_{i-1}}
  \label{eq:ptpvar}
\end{align}
in units of degrees per time unit. The mean of the point-to-point variation
\begin{align}
  m = \left< \left( \Delta \chi / \Delta t \right)_i \right>
\end{align}
indicates a secular trend in the data. For a linear increase or decrease of
the EVPA this value equals the slope of the linear regression.  A
point-to-point variation on the order of the trend indicates a smooth
variation; whereas a strong deviation from the trend indicates a
point-to-point variation larger than the general trend. The
point-wise deviation of the variation from the trend is calculated as
\begin{align}
	s_i = \left( \frac{\Delta \chi}{\Delta t} \right)_i - \left< \left( \frac{\Delta \chi}{\Delta t} \right)_i \right>.
\end{align}
The mean over this deviation is used as an estimator for the
smoothness of the EVPA curve with respect to a potential linear trend:
\begin{align}
  s = \left< s_i \right> = \left< \left( \frac{\Delta \chi}{\Delta t} \right)_i - \left< \left( \frac{\Delta \chi}{\Delta t} \right)_i \right> \right>.
  \label{eq:smoothness}
\end{align}
The smoothness estimator $s$ is the mean absolute deviation over the
mean (MAD) of the point-to-point variation as defined in
eq.~\ref{eq:ptpvar}. An EVPA curve with $s_1$ is considered
\emph{smoother} than a second curve with $s_2 > s_1$.

%_NEW_SUBSECTION________________________________________________________________
\subsection{Q-U-Random walk process}
\label{sec:randomwalk}

\begin{table*}[t]
\centering
\caption{Simulation results for the \textit{simple} and \textit{shock
    random walk process}: frequencies of EVPA rotation amplitudes,
  smoothness estimators, EVPA trends, their mean over all
  simulations, and the frequency of both shifting methods leading to
  the same EVPA curve.}
\label{tab:simresults}
{\small
\begin{tabular*}{\linewidth}{@{\extracolsep{\fill} } l r r r r}
	\hline \hline
		& \multicolumn{2}{c}{Simple random walk process}	& \multicolumn{2}{c}{Shock random walk process}	\\
	EVPA rotation amplitude $A_\chi$:	& Method~1:		& Method~2	& Method~1:		& Method~2		\\
	\hline
	$A_\chi > 180\,^\circ$:	& $99.50\,\%$		& $99.50\,\%$	& $99.75\,\%$		& $99.75\,\%$	\\
	$A_\chi > 360\,^\circ$:	& $42.49\,\%$		& $42.49\,\%$	& $50.5\,\%$		& $50.5\,\%$	\\
	$A_\chi > 720\,^\circ$:	& $0.41\,\%$		& $0.41\,\%$	& $0.90\,\%$		& $0.90\,\%$	\\
	\hline
	\multicolumn{3}{l}{smoothness estimator $s$:}					\\
	\hline
	$s < 6\,^\circ / \mathrm{d}$:	& $>0\,\%$	& $>0\,\%$	& $>0\,\%$		& $>0\,\%$	\\
	$s < 8\,^\circ / \mathrm{d}$:	& $0.008\,\%$	& $>0\,\%$	& $0.022\,\%$		& $>0\,\%$	\\
	$s < 10\,^\circ / \mathrm{d}$:	& $0.18\,\%$	& $0.44\,\%$	& $0.08\,\%$		& $0.22\,\%$	\\
	$s < 20\,^\circ / \mathrm{d}$:	& $79\,\%$	& $98\,\%$	& $78\,\%$		& $98\,\%$	\\
	$ \left< s \right> =$		& $17.5\,^\circ / \mathrm{d}$	& $15.1\,^\circ / \mathrm{d}$	& $17.7\,^\circ / \mathrm{d}$	& $15.1\,^\circ / \mathrm{d}$		\\
	\hline
	\multicolumn{3}{l}{EVPA trend $m$:}	\\
	\hline
	$|m|> 1\,^\circ / \mathrm{d}$	& $66\,\%$	& $66\,\%$	& $69\,\%$		& $69\,\%$	\\
	\hline \hline
	$\chi_\mathrm{sm1} = \chi_\mathrm{sm2}$:		& 	\multicolumn{2}{c}{$1.28\,\%$}	& \multicolumn{2}{c}{$0.87\,\%$}	\\
	\hline
\end{tabular*}
}
\end{table*}

To simulate a stochastic variation of the EVPA we perform a random
walk in Stokes-Q-U-plane.  The model contains $N$ cells of equal
intensity $I$.  The number $N$ determines the mean degree of linear
polarization
\begin{align}
	\left< P \right> = \sqrt{\frac{\pi}{4N}} \cdot P_\mathrm{max}
\end{align}
with $P_\mathrm{max} = 75\,\%$ \cite{2007ApJ...659L.107D}.

Each cell contains a uniform, but randomly oriented magnetic field.
We draw random samples from uniform distributions of Stokes $Q_i \in
[-1, +1]$ and $U_i \in [-1, +1]$ and normalize $Q_i$ and $U_i$ with
the factor $P_\mathrm{max}/\sqrt{Q_i^2 + U_i^2}$; with $i \in [1,
  N]$. The sums $Q = \sum_i^N Q_i$ and $U = \sum_i^N U_i$ determine
the integrated degree of linear polarization, $P$, and electric vector
position angle, $\chi$:
\begin{align}
	P &= \frac{\sqrt{Q^2 + U^2}}{I}	\\
	\chi &= \frac{1}{2} \arctan \frac{U}{Q} + n \frac{\pi}{2}	\\
	\text{with \ \ } n &=
	\begin{cases}
			1, & \text{if}\ Q<0 \\
			0, & \text{otherwise}
		\end{cases}
\end{align}

The variation of the polarization properties $P$ and $\chi$ is
determined by the number of cells $N_\mathrm{var}$ that change the
magnetic field orientation in each mean time step $\left< \Delta t
\right>$ \citep{2007ApJ...659L.107D}.  The fraction of cells
$X_{\left< \Delta t \right>}$ varying per mean time step $\left<
\Delta t \right>$ is estimated by the standard deviation $\sigma(P)$
of the degree of linear polarization:

\begin{align}
	\frac{N_\mathrm{var}}{N} = X_{\left< \Delta t \right>} \sim \frac{\sigma(P)}{\left< P \right>}.
\end{align}
For modeled time steps $\Delta t$ we scale the fraction of changing
cells linearly with the time step:
\begin{align}
	X_{\Delta t} &= \frac{\Delta t}{{\left< \Delta t \right>}} X_{\left< \Delta t \right>}		\\
	N_\mathrm{var}^{\Delta t} &= X_{\Delta t} \cdot N
	\text{	 rounded to integer}	\\
	\text{	with } N_\mathrm{var}^{\Delta t} &\leq N
\end{align}
We define two slightly different random walk processes.

\begin{description}
	\item[Simple random walk process:] Each time step
          $N_\mathrm{var}$ cells are selected randomly. The Stokes
          vectors of this subset of cells are reset randomly as
          described before.
	\item[Shock random walk process:] To simulate a shock passing
          through a turbulent jet, the cells are numbered. Each time
          step the cells $i = 1..N_\mathrm{var}$ are deleted and
          $N_\mathrm{var}$ new cells with randomized Stokes vectors
          are appended \cite[e.g.,][]{1985ApJ...290..627J,2007ApJ...659L.107D}.
\end{description}

The observed $\left< P \right>$ corresponds to $N = 31$.  The number
of cells changed each mean time step is $N_\mathrm{var} = x \cdot N =
21$.  Averaged over all simulations the resulting standard deviation
of the linear polarization degree is $\left< \sigma(P) \right> =
6\,\%$, smaller than the measured $8\,\%$.  Even with the maximum
number of cells changing, $N_\mathrm{var} = N$, the averaged standard
deviation of the linear polarization degree is $\left< \sigma(P)
\right> < 8\,\%$.  This already indicates that the polarization curve
of 3C\,279 is not produced by a stochastic process, at least not with
one having the probability density functions that are used in our
simulations.

The mean time step of our observations is $\left< \Delta t \right> =
3\,\mathrm{d}$.  The simulation time series is constructed randomly
with time steps $\Delta t$ following a power law distribution
$\mathrm{P}(\Delta t) \propto \Delta t^\alpha$ with $\alpha < -1$,
within the limits $[\Delta t_\mathrm{min}; \Delta t_\mathrm{max}]$.
This simulates a time step distribution similar to the observed data.
The EVPA is then calculated following either the \textit{simple random
  walk process} or the \textit{shock random walk process}.  Simulated
measurement errors are set randomly following a Gaussian distribution
with standard deviation, $\sigma_\mathrm{errors}$.  The resulting EVPA
is modified with both shifting methods.  For each modified EVPA curve
we determine the amplitude of variation, $A_\chi = \chi_\mathrm{max} -
\chi_\mathrm{min}$, and calculate the smoothness estimator, $s$, and
EVPA trend, $m$.

The simulation input parameters are shown in
Table~\ref{tab:simparameters}.  Large EVPA rotations ($> 180\,^\circ$)
have been observed at the time scale of days to 100\,days
\cite[e.g.,][]{2008A&A...492..389L, 2010Natur.463..919A,
  2010ApJ...710L.126M}.  We simulate time intervals of $T =
200\,\mathrm{d}$, to measure the frequency of large rotations within
this time scale.  We run the simulation $1\,000\,000$ times for both
random walk processes.

\begin{table}
\centering
\caption{Smoothness estimator $s$ (errors in parantheses) and shifting consistency of the EVPA of 3C\,279 in four observation epochs.}
\label{tab:3c279evpavar}
{\small
\begin{tabular*}{\columnwidth}{@{\extracolsep{\fill} } l l r r}
	\hline \hline
	epoch	& JD-2450000		& $s [\mathrm{^\circ/d}]$		& $\chi_\mathrm{sm1} = \chi_\mathrm{sm2}$	\\
	\hline
	I	& 5150 - 5310		& $32(5)$			        & no					\\
	II	& 5310 - 6050		& $5.0(4)$				& yes					\\
	IIb	& 5660 - 5760		& $4.4(5)$				& yes					\\
	III	& 6050 - 6110		& $10.5(8)$				& yes 				        \\
	\hline
\end{tabular*}
}
\end{table}

%_NEW_SUBSECTION________________________________________________________________
\subsection{Simulation results}
\label{sec:simresults}

Table~\ref{tab:simresults} lists the frequencies of EVPA amplitudes
$A_\chi$, smoothness estimators $s$, EVPA trends $m$ and the
corresponding means over all simulations for both random walk
processes and both EVPA shifting methods. The last row shows the
probability that both EVPA shifting methods give the same result.
The main simulation results are:
\begin{itemize}
	\item Large EVPA rotations of $> 360\,^\circ$ in less than
          $200\,\mathrm{d}$, as observed in our data, are common in
          random process based EVPA curves. The simple and the shock
          random walk processes produce rotations of that order with a
          probability of $43\,\%$ and $51\,\%$, respectively.
	\item Only in fewer than $1.3\,\%$ of the simulations do both
          EVPA shifting methods result in EVPA curves that are
          consistent with each other.
	\item A smoothness estimator $s < 8\,^\circ/\mathrm{d}$ occurs
          less than 30~ times in 1.000.000 simulations and $s$ is
          larger than $10\,^\circ/\mathrm{d}$ with a probability
          $>99.7\,\%$.
\end{itemize}

%_NEW_SECTION___________________________________________________________________
\section{EVPA variation in 3C\,279}
\label{sec:3C279evpa}

\begin{figure*}
\centering
\sidecaption
\includegraphics[width=0.72\textwidth,clip]{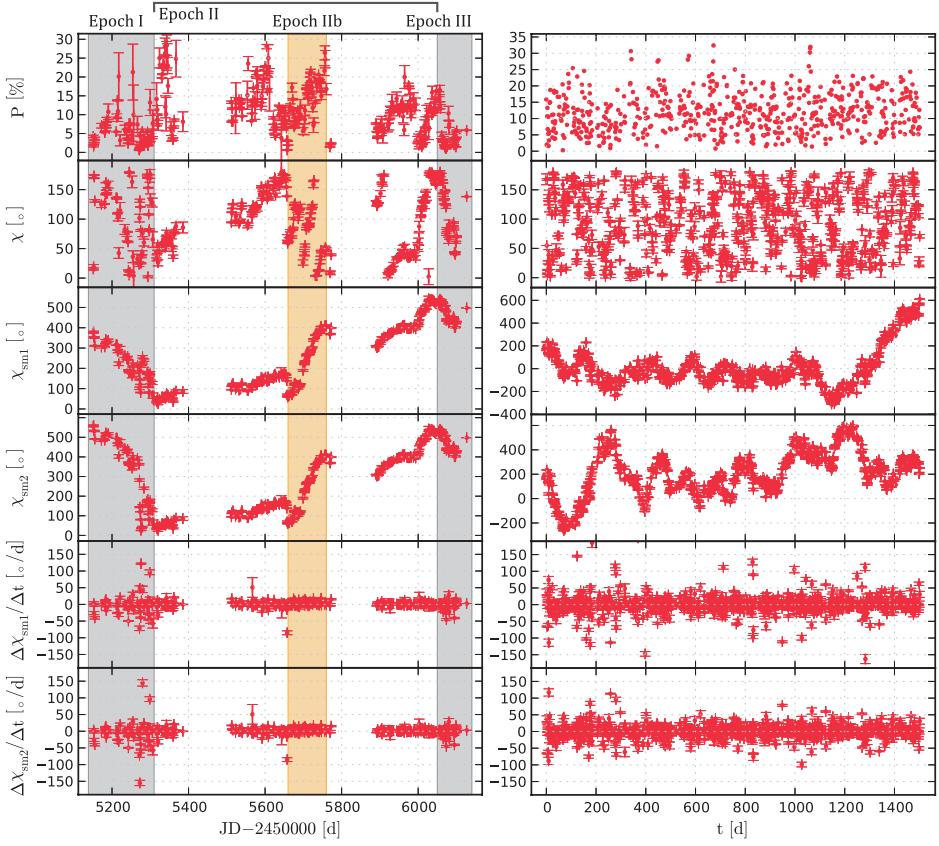}
\caption{Polarization variation of 3C\,279 (left plot) and a
  simple random-walk process simulation (right plot). The panels
  show from top to bottom: degree of linear polarization, EVPA in the
  $180\,^\circ$ interval, EVPA shifted with method~1, EVPA shifted
  with method~2, point-to-point variation (eq.~\ref{eq:ptpvar}) of the
  EVPA shifted with method~1 and method~2.}
\label{fig:EVPAvariation}
\end{figure*}

Figure~\ref{fig:EVPAvariation} shows a comparison of the measured
polarization variation of 3C\,279 (left plot) and one simulation over
the same time interval as the observation, based on the simple random
walk process (right plot).  We make four general observations: 1) The
observed polarization degree and EVPA ($180\,^\circ$ interval)
variation is less erratic than the simulated polarization degree and
EVPA variation.  2) This simulation shows an EVPA variation amplitude
of the same order as the observed, but 3) the simulated EVPA curve is
less smooth than the observed one. 4) The smoothness of the observed
EVPA curve is shown quantitatively in the small scatter of the
point-to-point variation (eq.~\ref{eq:ptpvar}) compared to the
simulation.  For a more detailed and quantitative description we
divide the observation time into four epochs.  Details are listed in
Table~\ref{tab:3c279evpavar}.

Epoch~I covers the major clockwise rotation of the EVPA. The
smoothness estimator is significantly higher than the mean smoothness
estimator of the random walk process simulations. The two EVPA
shifting methods give inconsistent results. Shifting method~1 gives a
total rotation of $\sim 350\,^\circ$, method~2 of $\sim 550\,^\circ$.
Both the high smoothness estimator and the shifting inconsistency
point to a possible random process origin of the EVPA variation during
epoch~I.

Epoch~II covers a general EVPA counter-clockwise rotation with
changing slope that includes one sharp EVPA decrease of $120\,^\circ$
in 4 days around JD2455655 and possibly a second decrease during the
observation gap around JD2455800.  The shifting results are consistent
and the smoothness estimator is $s = 5.0(4)$. Our simulation gives a
probability of $< 10^{-6}$ for creating EVPA variation with these
characteristics from a random process. Epoch~IIb is a subset of
epoch~II that shows a smooth ($s=4.4(5)$), counter-clockwise EVPA
rotation of $360\,^\circ$ in $\sim 100$ days. During epoch~III the
EVPA decreases. The shifting methods are consistent, but the
smoothness estimator increased with respect to the previous epoch.

This analysis indicates that two different processes affect the EVPA
variation.  During epoch~I the jet is in a low state ($F_\nu <
1\,\mathrm{mJy}$ (R-band)) and the EVPA variation likely originates
from random fluctuations of the net B-field orientation in a turbulent
emission region. At the end of epoch~I - around JD2455350 - the degree
of linear polarization increases by a factor of $2 - 5$.  A new
emission feature likely emerges in the jet.  This new feature
dominates the jet emission, increasing by a factor of $\sim 20$ from
epoch~1 to the flare peak, and causes a deterministic EVPA rotation
coinciding with flaring epoch~2.  Different processes can explain
a deterministic EVPA variation:
1)~Abdo~et.~al.~\cite{2010Natur.463..919A} explained the EVPA rotation
of 3C\,279 in February~2009 with a globally bending jet. Bends in
different directions could explain the two-directional EVPA swings,
but a rotation of at least $360\,^\circ$ requires the jet to follow a
helical path.  2)~Two orthogonally polarized emission features shift
the integrated EVPA, when one of the features fades in or out.  This
process can rotate the EVPA in both directions, but only with
amplitudes $< 90\,^\circ$ \cite{1984MNRAS.211..497H}.  3)~Two models
are based on a jet emission feature - possibly a moving shock - that
does not fill the full jet cross-section and moves along a helical or
bent streamline.  In one of them, the magnetic field can be tangled
and the moving shock compresses and partially orders the tangled
magnetic field \cite{1990ApJ...350..536C}. The rotation of the EVPA is
then coupled to the bent motion of the shock front
\cite{2010IJMPD..19..701N, 2013ApJ...768...40L}.  In the other one the
magnetic field is ordered - either helical or toroidal - and the
feature `highlights' different parts of the magnetic field, resulting
in an EVPA swing \cite{1988A&A...190L...8K, 2008Natur.452..966M}.
Both of the latter models are expected to produce uni-directional EVPA
swings only, as the rotation is coupled to the motion of the feature
within the jet cross-section.  If the emission feature is ejected from
the accretion disk the twist of the feature originates in the angular
velocity of the disk and is not expected to change direction.
However, our first modeling results - in preparation - indicate that
two-directional EVPA rotations could indeed be reproduced in the
framework of an emission feature following a helical trajectory in a
helical magnetic field, assuming angular momentum conservation in a
jet that is opening up downstream.

During epoch~III the flux decreases, suggesting that the emission feature
fades out and the random variations in the low state jet start to
dominate again, consistent with the increase of the smoothness
estimator.

%_NEW_SECTION___________________________________________________________________
\section{Conclusions}
\label{sec:conclusions}

When `solving' the $n \pi$-ambiguity, one faces the danger of wrongly
reconstructing the actual EVPA curve and misinterpreting the
data. Poor sampling rates may obscure large rotations. And shifting
data points by multiple times of $\pi$ may artificially introduce
large rotations that actually are random and non-directed
variations.

We present a method to distinguish between possibly random EVPA
variation and deterministic EVPA variation. We run Monte Carlo
simulations of a random walk in Stokes-Q-U-space, adjusted to
reproduce the mean and standard variation of the linear polarization
degree that we observe, with a randomized sampling rate, that follows
the same characteristics as our data.  The introduced smoothness
estimator quantifies the smoothness of the curve.  The simulations of
different kinds of random walk processes give us a threshold for the
smoothness estimator. We can exclude these models at a certain
confidence level, if the smoothness estimator of the observed data is
smaller than the threshold. We point out that the threshold depends on
the data sampling characteristics and the number of modeled cells, and
thus the quoted values are only valid for our data set. The smoothness
estimator threshold and the (in)consistency of the two presented
shifting methods are two indicators to distinguish between possibly
random and deterministic EVPA variation.

3C\,279 shows both characteristics in different states.  In the low
state the EVPA rotation is consistent with a stochastic process,
possibly due to turbulence in the jet. In the flaring state the EVPA
behaviour is likely deterministic and possibly dominated by a single
emission feature. We exclude the model of a bent jet and the simple
two-component model. A helical trajectory of the emission feature
seems to be able to explain the EVPA rotation and modeling the EVPA
curve in this scenario is in progress. Also, modeling will be needed
to discriminate between the two possibilities of a tangled and shock
compressed magnetic field or a structured (toroidal or helical)
field.

Optical flux density and EVPA during the $360\,^\circ$ rotation
(epoch~IIb) show a high resemblance to the EVPA variation of
PKS\,1510$-$089 in March~2009 and BL~Lac in September~2005, all
coinciding with a general optical flaring episode and ending with
sharp sub-flare \cite{2010ApJ...710L.126M, 2008Natur.452..966M}. Thus,
we observe the flare-related EVPA rotations not only in different
objects, but in different classes of objects (quasars and
BL~Lacs). The $360\,^\circ$ rotation in 3C\,279 takes
$110\,\mathrm{d}$.  Assuming a Lorentz factor $\Gamma = 15$, this time
corresponds to a traveled distance of $\Delta r \approx 5 \cdot 10^5$
Schwarzschild radii (assuming a BH mass of $6\times10^8 M_\odot$
\cite[and references therein]{2010Natur.463..919A}). If this emission
feature is located in the acceleration and collimation zone, the mean
Lorentz factor and traveled distance are likely smaller.

With an increasing interest in well-sampled polarization data
\cite[e.g.,][]{2008Natur.452..966M, 2010ApJ...710L.126M} and new
optical polarimetry monitoring projects like
\textsc{Robopol} (\url{http://robopol.org/}), our analysis
method may prove to be useful in discriminating between stochastic and
deterministic EVPA rotations. As we observe two different processes in
the low state and the flaring state of 3C\,279, we point out that it
is mandatory not only to observe blazars triggered during flaring
states, but also with good sampling during low states to fully understand
the structure of the magnetic field and the processes responsible for
EVPA variation.

%ACKNOWLEDGEMENTS%%%%%%%%%%%%%%%%%%%%%%%%%%%%%%%%%%%%%%%%%%%%%%%%%%%%%%%%%%%%%%%

\begin{acknowledgement}
	S.K. was supported for this research through a stipend from
        the International Max Planck Research School (IMPRS) for
        Astronomy and Astrophysics at the Max Planck Institute for
        Radio Astronomy in cooperation with the Universities of Bonn
        and Cologne. The research at Boston University was partly
        funded by NASA Fermi GI grant NNX11AQ03G. K.V.S. and
        Y.Y.K. are partly supported by the Russian Foundation for
        Basic Research grant 13-02-12103. N.G.B. was supported by the
        RFBR grant 12-02-01237a. St.~Petersburg University team was
        supported by the RFBR grants 12-02-00452 and
        12-02-31193. E.B., M.S. and D.H. thank financial support from
        UNAM DGAPA-PAPIIT through grant IN116211-3. The research at
        the IAA-CSIC is supported by the Spanish Ministry of Economy
        and Competitiveness and the Regional Government of Andaluc\'ia
        (Spain) through grants AYA2010-14844 and P09-FQM-4784,
        respectively. The Calar Alto Observatory is jointly operated
        by the Max-Planck-Institut f\"ur Astronomie and the Instituto
        de Astrof\'isica de Andaluc\'ia-CSIC. Data from the Steward
        Observatory spectropolarimetric monitoring project were
        used. This program is supported by Fermi Guest Investigator
        grants NNX08AW56G, NNX09AU10G, and NNX12AO93G. We acknowledge
        the photometric observations from the AAVSO International
        Database contributed by observers worldwide and used in this
        research. This paper has made use of up-to-date SMARTS
        optical/near-infrared light curves that are available at
        \url{http://www.astro.yale.edu/smarts/glast/}
        \cite{2012ApJ...756...13B}.
\end{acknowledgement}

%BIBLIOGRAPHY%%%%%%%%%%%%%%%%%%%%%%%%%%%%%%%%%%%%%%%%%%%%%%%%%%%%%%%%%%%%%%%%%%%

\bibliography{references}

\end{document}